\documentclass[conference]{IEEEtran}
\IEEEoverridecommandlockouts
\usepackage{cite}
\usepackage{amsmath,amssymb,amsfonts}
\usepackage{algorithmic}
\usepackage{graphicx}
\usepackage{textcomp}
\usepackage{xcolor}
\def\BibTeX{{\rm B\kern-.05em{\sc i\kern-.025em b}\kern-.08em
    T\kern-.1667em\lower.7ex\hbox{E}\kern-.125emX}}
\usepackage{amsmath,amssymb,mathrsfs}
\usepackage{amsthm}
\usepackage{epsfig,epsf,graphicx,graphics}
\usepackage{url}
\usepackage{array}
\usepackage{multicol}
\usepackage{empheq}
\usepackage[caption=false, font=footnotesize]{subfig}

\newcommand{\lp}{\left(}
\newcommand{\rp}{\right)}
\newcommand{\lb}{\left[}
\newcommand{\rb}{\right]}
\newcommand{\mv}{\middle\vert}

\newcommand{\ol}{\overline}
\newcommand{\mcal}{\mathcal}

\newcommand{\mrm}{\mathrm}

\newcommand{\mb}{\mathbf}

\allowdisplaybreaks

\begin{document}

\title{Modeling Friendship Networks among Agents with Personality Traits
}

\author{\IEEEauthorblockN{
Heng-Chien Liou}
\IEEEauthorblockA{\textit{Department of Electrical Engineering} \\
\textit{National Taiwan University}\\
Taipei, Taiwan \\
liouhengchien@gmail.com}
\and
\IEEEauthorblockN{
Hung-Yun Hsieh}
\IEEEauthorblockA{\textit{Graduate Institute of Communication Engineering} \\
\textit{National Taiwan University}\\
Taipei, Taiwan \\
hungyun@ntu.edu.tw}
}

\maketitle

\begin{abstract}

Using network analysis, psychologists have already found the nontrivial correlation between personality and social network structure. Despite the large amount of empirical studies, theoretical analysis and formal models behind such relationship are still lacking. To bridge this gap, we propose a generative model for friendship networks based on personality traits. To the best of our knowledge, this is the first work to explicitly introduce the concept of personality and friendship development into a social network model, with supporting insights from social and personality psychology. We use the model to investigate the effect of two personality traits, extraversion and agreeableness, on network structure. Analytical and simulation results both concur with recent empirical evidence that extraversion and agreeableness are positively correlated with degree. Using this model, we show that the effect of personality on friendship development can amount to the effect of personality on friendship network structure. 
\end{abstract}


\section{Introduction}

Over the past few decades, network analysis in the social science has experienced explosive growth, and generated tremendous amount of researches and corresponding insights \cite{burt2013social}. Its increased popularity in psychology have stimulated the attempt to merge the structural approaches of network analysis with the individual differences literature \cite{selden2018review}. 

One of the fundamental topics is the relationship between personality and network structure \cite{fang2015integrating, selden2018review}. For example, research has found that extraversion and agreeableness relate consistently to personal networks, and openness predicts network diversity \cite{selden2018review}. Most of these works are correlational studies, collecting data from pre-existing social networks, and utilize statistical tools to find the correlation between variables. These studies provide the empirical ground for the relationship between personality and network structure. However, the theoretical insight behind such relationship is still lacking. Specifically, there is no formal model to describe how the relationship is generated in current literature. 

Focusing on friendship networks, in this study, we propose a generative network model for friendship networks with personality. We theorize that the effect of personality on friendship development can amount to the effect of personality on network structure. We further apply the model to investigate two specific personality traits, extraversion and agreeableness. The results show that both traits are positively correlated with degree, a finding supported by empirical evidence, thus validating the applicability of the proposed model. 

The rest of paper is organized as follows. In Section \ref{related}, we review the empirical research and formal models about personality and friendship networks. In Section \ref{fri_related}, we introduce the concepts and findings from science of personality and friendship that inspired our model. We then propose and analyze the model in Section \ref{model}. After considering practical modeling scenario in Section \ref{modeling}, the results are validated in Section \ref{validation}. We discuss the implication and outlooks in Section \ref{discussion} and conclude in Section \ref{conclusion}. 
\section{Related Work}
\label{related}

Focusing on personality and friendship networks, in this section, we first review recent findings from empirical studies. And then we discuss why existing models are not sufficient to explain the phenomena observed. 

\subsection{Personality and Friendship Networks in Empirical Research}

While there is a large amount of literature examining how personality and social networks are intertwined, we focus on representative ones to gain more insights.  
\cite{fang2015integrating} used data from 138 independent samples to conduct meta-analysis, finding that self-monitoring predicts in-degree centrality in both expressive and instrumental networks. 
\cite{selden2018review} reviewed 30 articles, concluding that extraversion and agreeableness are consistently related to personal but not workplace networks, openness is predictive of network diversity, and conscientiousness is associated with maintaining certain personal networks.

Specifically, there is also literature that explicitly focused on friendship networks. For example, \cite{feiler2015popularity} used the friendship network of an incoming cohort of students, showing that extraverts accumulate more friends than introverts do; \cite{selfhout2010emerging} collected sociometric nominations and self‐ratings on personality traits during the first year of university, concluding that an individual who is high on extraversion has higher out-degree, and an individual who is high on agreeableness has higher in-degree. Combining with other studies \cite{pollet2011extraverts, zhu2013pathways, wagner2014belongs, kalish2006psychological, schulte2012coevolution}, in short, we can conclude that both extraversion and agreeableness are correlated with degree in friendship networks. This finding is also consistent with the result for general social networks summarized by \cite{selden2018review}. 

\subsection{Model of Friendship Networks and Models with Personality}

There have been several works addressing the model of friendship networks. \cite{singer2009agent} proposed an  agent-based model for structuring of friendship networks in a campus-like fixed setting, with friendship formation based on the frequency of encounters and mutual interest, to produce self-organized community structures. \cite{currarini2009economic} proposed a game-theoretic framework for friendship to understand the racial segregation in friendship networks. While there is plenty of research concerning friendship networks, none of them has incorporated personality. 

To the best of our knowledge, the only generative model for social networks with personality was proposed by \cite{ilany2016personality}. However, the model was proposed under the context of animal social networks with only one personality considered, namely boldness. Human social networks are expected to be different from animal social networks \cite{lusseau2004identifying}, and boldness is not even included as primary trait in human psychology \cite{mccrae1992introduction}. 

In conclusion, to the best of our knowledge, there are no existing models suitable to study human friendship networks with personality. Intended to bridge this gap and explain the observed correlation between personality and network structure theoretically, we propose a generative model and show that the microscopic difference in friendship formation can amount to the observed results. 
\section{Science of Personality and Friendship}
\label{fri_related}

Personality and friendship have been two persistent topics of interest throughout history. Both of them have stimulated profound discoveries about human life and established multifarious schools of thoughts. While these two topics are studied separately most of the time, the interdependence has been investigated only recently \cite{harris2016friendship}.

\subsection{Personality}

Within psychology, personality is a set of psychological traits, which contribute to the enduring and distinctive patterns of feelings, thoughts, and behaviors of individuals \cite{cervone2015personality}. Among various paths toward the understanding of personality, the Five Factor Model (FFM) is a widely accepted model, using five orthogonal dimensions to explain or even predict individuals' feelings, thoughts and behaviors \cite{mccrae1992introduction}. The Five Factors or the Big Five traits are \textit{Openness}, \textit{Conscientiousness}, \textit{Extraversion}, \textit{Agreeableness}, \textit{Neuroticism} \cite{mccrae1992introduction}. 

\begin{figure}[!t]
    \centering
    \includegraphics[width=70mm]{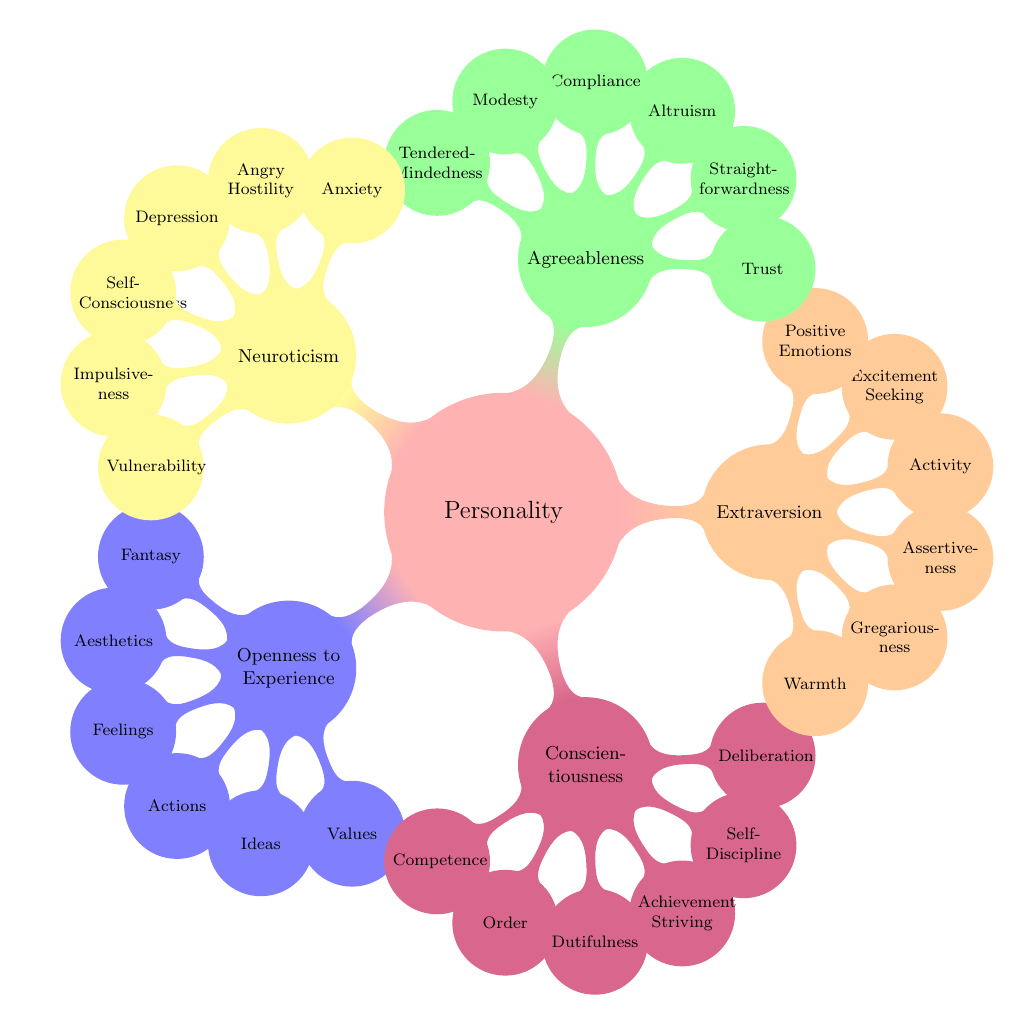}
    \caption{Visual illustration of the hierarchy of Five Factor Model. The image is drawn according to Costa and McCrae's assessment \cite{costa1995domains}. }
    \label{bigfive_fig}
\end{figure}

Openness (O) describes the tendency of having wide interests and being imaginative, curious, and artistic \cite{mccrae1992introduction}. Conscientiousness (C) is considered to be a dimension that holds impulsive behaviors in check, organizes the behaviors, and makes people behave thorough, well-organized, and achievement oriented \cite{mccrae1992introduction}. Extraversion (E) is characterized by being cheerful, talkative, sociable, and motivated to engage in interpersonal interaction \cite{costa1995domains, mcadams2015art}. Agreeableness (A) is a dimension about being warm, caring, altruistic, and easy to go along with \cite{costa1995domains}. Neuroticism (N) represents the tendency to experience distress and negative affect, and in the cognitive and behavioral style following \cite{mccrae1992introduction, costa1995domains}. 

Beside the above description, the five traits can also be evaluated more deliberately by using more specific constructs, facets. The hierarchy of personality trait, including five domains and the corresponding facets, is shown in Fig. \ref{bigfive_fig}. 

\subsection{Friendship}

With the definition proposed by Dunbar, friends are the people who share the lives in a way that is more than just the casual meeting of strangers \cite{dunbar2018anatomy}. 
One important thing to note is that friendship is not static. Instead, friendship is dynamic and changes over times \cite{adams1994integrative}. The course of friendship change can be roughly summarized by the following phases: \textit{formation}, \textit{maintenance}, and possibly \textit{dissolution} \cite{adams1994integrative}.

At the beginning of friendship, the formation phase, two strangers become acquainted to each others, through either face-to-face encounter or other means. The two individuals develop connection and identify each other as friend. 

After the two formed the relationship, the course of friendship enters into another phase. In the maintenance phase, people adopted different ways to sustain their interest and involvement. Conflicts may arise because of various external and internal factors, which require volition to reconcile. In some relationship, the people involved fail to reconcile the conflicts or loss their interests in maintaining the connection, which leads to the dissolution of friendship. 

\subsection{Interdependence of Personality and Friendship}

As already shown, many facets in personality reflect the individual difference in social behavior. It is not surprising to find out that personality and friendship are intertwined. 

Generally, extravesion and agreeableness both have a positive effect on friendship development \cite{harris2016friendship}. 
{Extraversion} is associated to being liked more at low level of acquaintance, more comfortable when interacting with strangers, and believe that they are likable \cite{cuperman2009big}. Agreeable people are better liked than the disagreeable after interactions \cite{cuperman2009big}; when there is at least one person in relationship high on {agreeableness}, both are more likely to feel comfortable and use constructive strategies to reconcile conflicts \cite{cuperman2009big}. 

Research suggested that being emotionally unstable is problematic in friendship maintenance \cite{harris2016friendship}. 
People high on neuroticism anticipate negative outcome in maintenance, and feel less secure. Neuroticism is associated with excessive reassurance-seeking, which might lead to  relationship degradation \cite{joiner1999depression}. Compared to other three factors, there is little evidence concerning the roles played by conscientiousness and openness to experience in friendship \cite{harris2016friendship}.

\section{Generative Model for Friendship Networks with Personality}
\label{model}


We propose a generative network model for friendship networks with personality. The governing rules in this model are formulated based on the discovery from psychology of personality and friendship: how each agent forms or dissolves friendship will depend on one's own personality and other's. The proposed model can be viewed as an extension of growing network models with heterogeneity \cite{ferretti2012features}. 

\subsection{Model Formulation}

\begin{figure}[!t]
    \centering
    \includegraphics[width=\linewidth]{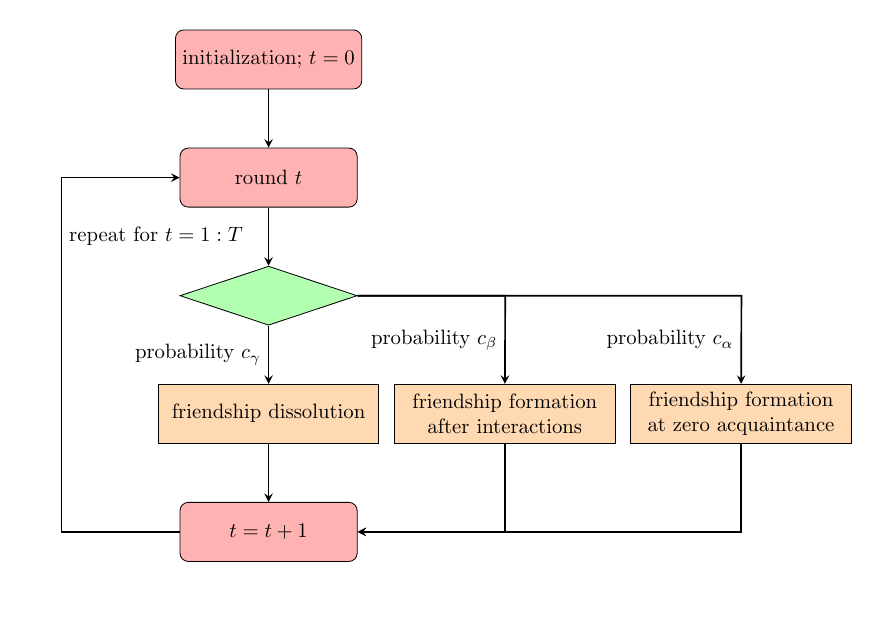}
    \caption{Visual illustration of the high level description of the proposed model. }
    \label{fig:highlevel}
\end{figure}

The formulated model will generate a sequence of undirected simple networks. Starting from an initial network, in each round, the model will do one of the three subroutines: adding a new node and forming edges from the new node ($\alpha$), forming edges between existing nodes ($\beta$), and dissolving edges between existing nodes ($\gamma$). Fig. \ref{fig:highlevel} illustrate the high level description of the proposed model. 

The probability distribution governing which edge to form or dissolve is determined by agents' personality and their degree. Formally, the following elements define the models.  

\begin{enumerate}
    \item $\mcal{P}$ is an arbitrary space. The elements $p$ are referred as the agents' \textit{personality}. 
    \item $\rho$ is a real non-negative function with unit measure over $\mcal{P}$, which is the probability distribution of personality in the corresponding population. 
    \item $c_\alpha$, $c_\beta$, and $c_\gamma$ indicates the probability that subroutine $\alpha$, $\beta$, and $\gamma$ occur at each round. 
    \item  $\pi_\alpha$, $\pi_\beta$ , and $\pi_\gamma$ are
    $\lp \mcal{P} \times \mathbb{N} \rp^2 \rightarrow \mb{R_+}$ functions
    . These functions are used to describe how the personality of the engaged individual affect their difference in friendship formation and maintenance. 
    \item  $m_\alpha$, $m_\beta$, and $m_\gamma$ are 
    $\mcal{P} \rightarrow \mathbb{N}$ functions.
    Same as previous elements, these functions are used to model agents' difference in friendship development. 
\end{enumerate}

In the generated undirected simple graph $\mcal{G}^{ \lp t \rp} = \lp \mcal{V}^{ \lp t \rp}, \mcal{E}^{ \lp t \rp} \rp$ after round $t$, each node $v_i \in  \mcal{V}^{ \lp t \rp} $ is associated with personality $p_i \in \mcal{P}$, assumed to be fixed during the evolution of model. We sometimes overload the notation $v_i^{ \lp t \rp}$ to denote the tuple $\lp p_i, k_i \lp t \rp \rp$, where $ k_i \lp t \rp$ is the degree of $v_i$ at the end of round $t$. The network evolution is governed by the following rules. 

\begin{figure}[!t]
    \centering
    \includegraphics[width=\linewidth]{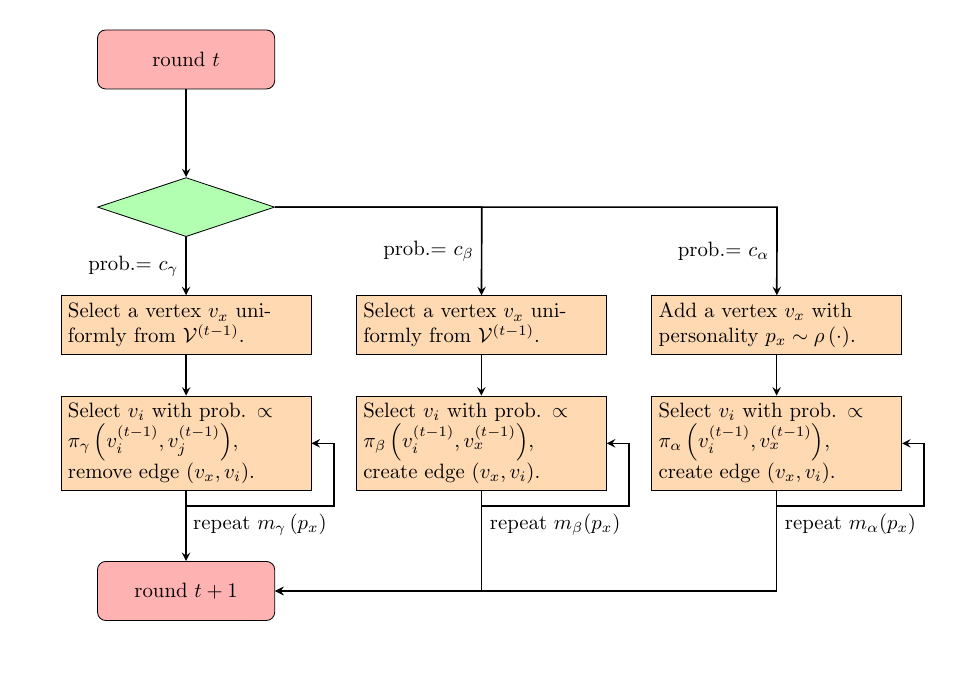}
    \caption{Visual illustration for modeling process at round $t$, where we use ``prob.'' as a shorthand of ``probability'', and ``$\propto$'' as a shorthand of ``proportional to''. }
    \label{model_fig}
\end{figure}

\begin{enumerate}
    \item The process starts with initial state (at round $0$) composed of $N_0$ agents and $L_0$ links, with arbitrary personality $p_i \in \mcal{P}$. 
    \item At each round $t$ (from $1$ to $T$), models will enter into three possible subroutines, with probability $\lp c_\alpha, c_\beta, c_\gamma \rp$:
    
    \textbf{Subroutine $\boldsymbol\alpha$: agent addition and friendship formation at zero acquaintance}
    
    With probability $c_\alpha$, a new agent $v_x$ with $m_\alpha(p_x)$ links attached is added to network. The newly added agent is randomly assigned a personality configuration $p_x$ according to distribution $\rho$. 
    
    Repeated for $m(p_x)$ iterations, a node $v_i$ in $\mcal{V}^{\lp t-1 \rp}$ is selected in each iteration with probability 
    proportional to $\pi_\alpha \lp v_i^{ \lp t-1 \rp}, v_x^{ \lp t-1 \rp}\rp$,
    and the edge $\lp v_x, v_i \rp$ is formed. 
    This subroutine models that when newcomers enters into a society, they may form friendship with others at zero acquaintance. 
    
    \textbf{Subroutine $\boldsymbol\beta$: friendship formation after interactions} 
    
    With probability $c_\beta$, an existing node $v_x \in  \mcal{V}^{\lp t-1 \rp}$ will be selected uniformly at random. 
    
    Repeated for $m_\beta(p_x)$ iterations, a node $v_i$ in $\ol{\mcal{N}^{\lp t-1 \rp} \lp v_x \rp} = \mcal{V}^{\lp t-1 \rp} - \mcal{N}^{\lp t-1 \rp} \lp v_x \rp$ is randomly selected in each iteration, with probability 
    proportional to $\pi_{\beta} \lp v_i^{ \lp t-1 \rp}, v_x^{ \lp t-1 \rp} \rp$
    , and the edge $\lp v_x, v_i \rp$ is created. 
    This subroutine models that for agents already existing in the society, they might still make friends with acquainted others. 
    
    \textbf{Subroutine $\boldsymbol\gamma$: friendship dissolution} 
    
    With probability $c_\gamma = 1- c_\alpha-c_\beta$, an existing node $v_x \in \mcal{V}^{\lp t-1 \rp} $ will be selected uniformly at random. 
    
    Repeated for $m_{\gamma} \lp p_x \rp$ iterations, an node $v_i$ in $\mcal{N}^{\lp t-1 \rp} \lp v_x \rp$, the neighborhood of $v_x$, is randomly selected in each iteration, with probability 
    proportional to $\pi_{\gamma} \lp v_i^{ \lp t-1 \rp}, v_x^{ \lp t-1 \rp} \rp$, 
    and the selected edge $\lp v_i, v_j \rp$ is removed. This subroutines models the potential friendship dissolution: after the friendship is formed, friendship might dissolve upon failures in maintenance. 
    
\end{enumerate}

Fig. \ref{model_fig} shows the modeling process in each round $t$, and the corresponding subroutines.

Note that the three subroutines do not directly correspond to the three phases in friendship development \cite{adams1994integrative}. We use Subroutine $\alpha$ and Subroutine $\beta$ to model friendship formation in two different scenarios; by doing so, we can distinguish the characterization of likability of people at zero acquaintance and after interactions \cite{back2011closer, cuperman2009big}; we use Subroutine $\gamma$ to model the dissolution upon failure in maintenance; by doing so, we can model the difference in conflict management during maintenance \cite{harris2016friendship}. 

For the ease of further analysis, we can relax the model to undirected weighted graph. The relaxed model is defined by the same elements as in original model, with matrix $W^{ \lp t \rp}$ relacing $\mcal{E}^{\lp t \rp}$. The matrix  $W^{ \lp t \rp}$ records the edge's weight between arbitrary two nodes after round $t$. With this adaption, between all pairs of node, the corresponding weights are initialized to $0$. Degree is defined as the sum of associating weights. Edge addition and edge removal becomes the incremental and decremental of corresponding weights. All procedures parallel the ones specified before, with $\mcal{V}^{\lp t-1 \rp}$ replacing both $ \ol{\mcal{N}^{\lp t-1 \rp} \lp v_x \rp}$ and $ {\mcal{N}^{\lp t-1 \rp} \lp v_x \rp}$. 


\subsection{Analysis}
\label{analysis}

To grasp the behavior of proposed model, we analytically derive the equilibrium. Since the stochastic model is hard to analyze, we adopt mean-field approximation from statistical physics \cite{ferretti2012features}. The mean-field rate equation goes as

\begin{align}
    & \quad N_k\lp p, t+1 \rp  \label{rate_eq} \\ 
    & = N_k \lp p, t \rp + \delta_{k, m_\alpha \lp p \rp} c_\alpha \rho \lp p \rp 
     + c_\alpha \Delta_\alpha + c_\beta \Delta_\beta + c_\gamma \Delta_\gamma, \nonumber 
\end{align}
where $N_k\lp p, t \rp$ denotes the expected number (or expected density) of node with degree $k$ and personality $p$ right after round $t$; $\Delta_\alpha$, $\Delta_\beta$, and $\Delta_\gamma$ denote the expected effects from three subroutines; $\delta_{k, m_\alpha \lp p \rp} $ is a Kronecker delta function.


\begin{table}[!t]
    \centering
    \caption{symbols and notations}
    \label{tab:symbols}
    \begin{tabular}{|l|l|}
        \hline
        $\mcal{P}$ & space for personality \\
        \hline 
        $\rho$ & probability distribution of personality \\
        \hline
        $c_\alpha, c_\beta, c_\gamma$ & {probability for each subroutines to occur} \\
        \hline 
        $\pi_\alpha$, $\pi_\beta$, $\pi_\gamma$ & {personality-degree-dependent preference function}  \\
        \hline
        $m_\alpha, m_\beta, m_\gamma$ & {function for number of edges added or deleted}  \\
        \hline
        $p$ (or $q$) & personality \\
        \hline
        $k$ (or $l$) & degree \\
        \hline 
        $N_k \lp p, t \rp$ & {expected number (density) of node with $\lp k,p \rp$} \\
        \hline 
        $\delta_{k, m_\alpha \lp p \rp} $ & {Kronecker delta: $\delta_{k, m_\alpha \lp p \rp}=1 \Leftrightarrow k = m_\alpha \lp p \rp$} \\
        \hline
        $\Delta_\alpha, \Delta_\beta, \Delta_\gamma$ & {expected change on $N_k \lp p, t \rp$ from subroutines} \\
        \hline
    \end{tabular}
\end{table}

We summarize the symbols and notation used in Table \ref{tab:symbols}. To illustrate how to analyze the model, we begin with some specific cases. 

\subsubsection{Edge changes come only from newcomers and are independent of degree}
\label{simplest}
We consider a simplest case for analysis by setting $c_\beta=c_\gamma = 0$, that is, friendship formation after interactions and dissolution never occur. 


We assume that $\pi_\alpha$ is independent of degree,

\begin{align}
    \pi_\alpha \lp \lp p, k \rp, \lp q, l \rp \rp
    = \sigma_\alpha \lp p, q \rp, \, \forall k, l \in  \mathbb{N}, \nonumber
\end{align}
that is, we drop the arguments of degree and preserve only the arguments for personality. 

Then we have 

\begin{align}
    \Delta_\alpha &= \sum_{q \in \mcal{P}} \frac{\sigma_\alpha \lp p, q \rp \lb N_{k-1} \lp p,t \rp - N_k \lp p, t \rp \rb }{\sum_{r \in \mcal{P}} \sigma_\alpha \lp r, q \rp \sum_{l \in \mb{N}} N_l \lp r, t \rp} \rho\lp q \rp m_\alpha \lp q \rp \nonumber \\
    &=  \frac{N_{k-1} \lp p,t \rp - N_k \lp p, t \rp}{c_\alpha t} \sum_{q \in \mcal{P}} \frac{\sigma_\alpha \lp p, q \rp \rho  \lp q \rp m_\alpha \lp q \rp}{\sum_{r \in \mcal{P}} \sigma_\alpha \lp r, q \rp \rho \lp r \rp}.  \label{alpha_simple}
\end{align}

For simplicity, let  

\begin{align}
    A \lp p \rp = \sum_{q \in \mcal{P}} \frac{\sigma_\alpha \lp p, q \rp  \rho  \lp q \rp m_\alpha \lp q \rp}{\sum_{r \in \mcal{P}} \sigma_\alpha \lp r, q \rp \rho \lp r \rp},
    \label{simplify_alpha}
\end{align}

Plugging (\ref{alpha_simple}) into (\ref{rate_eq}) with $c_\beta=c_\gamma = 0$, we get 

\begin{align}
    N_k\lp p, t+1 \rp &= N_k \lp p, t \rp + \delta_{k, m_\alpha \lp p \rp} c_\alpha \rho \lp p \rp \nonumber \\
     & \quad + \lp \frac{N_{k-1} \lp p,t \rp - N_k \lp p, t \rp}{ t} \rp A \lp p \rp. \nonumber
\end{align}

Substituting the normalized $n_k \lp p, t \rp = \frac{N_k \lp p, t \rp}{c_\alpha t} $, which is the degree-personality joint distribution, we have

\begin{align}
    n_k\lp p, t+1 \rp \lp 1+\frac{1}{t} \rp &= n_k \lp p, t \rp + \frac{\delta_{k, m_\alpha \lp p \rp} \rho \lp p \rp}{t} \nonumber \\
     & \quad + \lp \frac{n_{k-1} \lp p,t \rp - n_k \lp p, t \rp}{ t} \rp  A \lp p \rp.  \nonumber
\end{align}

For thermodynamic limit $n_{k} \lp  p  \rp = \lim_{t\rightarrow \infty} n_{k} \lp  p , t \rp$, the rate equation becomes 

\begin{align}
    n_k\lp p \rp = \lp n_{k-1} \lp p \rp - n_k \lp p \rp \rp  A \lp p \rp + \delta_{k, m_\alpha \lp p \rp} \rho \lp p \rp. \nonumber
\end{align}


Solving the recurrence subjected to boundary condition, 
 
\begin{align}
    n_k \lp p \rp = \frac{\rho \lp p \rp}{1 + A \lp p \rp} \lp \frac{A \lp p \rp}{1 + A \lp p \rp} \rp^{k-m_\alpha \lp p \rp}.
    \label{simplest-solution}
\end{align}

Note that we can also write down degree distribution conditioned on personality as

\begin{align}
    n \lp k \mv p \rp = \frac{n_k \lp p \rp}{\rho \lp p \rp} =  \frac{ 1}{1 + A \lp p \rp} \lp \frac{A \lp p \rp}{1 + A \lp p \rp} \rp^{k-m_\alpha \lp p \rp}, \nonumber
\end{align}
which is a shifted geometric distribution. 

Hence, we can write down the expected degree conditioned on personality, 

\begin{align}
    \mb{E} \lb k|p \rb = A \lp p \rp + m_\alpha \lp p \rp.
    \label{conditional-exp_simplest}
\end{align}

\subsubsection{Edge changes come from all three subroutines and independent of degree}

We consider a more complicated case build on the foundation in Section \ref{simplest}. 

We assume that $\pi_\alpha$, $\pi_\beta$, and $\pi_\gamma$ are all independent of degree, replaced by $\sigma_\alpha$, $\sigma_\beta$, and $\sigma_\gamma$ with domain $\mcal{P}^2$, dropped the arguments of degree. 


We proceed further by considering the model on weighted graph. $\Delta_\alpha$ is the same as in (\ref{alpha_simple}), while 

\begin{align}
    \Delta_\beta &=  \frac{N_{k-1} \lp p,t \rp - N_k \lp p, t \rp}{c_\alpha t} B \lp p \rp \label{beta_simple}, \\
    \Delta_\gamma &=  \frac{N_{k} \lp p,t \rp - N_{k-1} \lp p, t \rp}{c_\alpha t} \lp - \Gamma \lp p \rp \rp, \label{gamma_simple}
\end{align}
where we simplify the expression by letting
\begin{align}
    B \lp p \rp &= \sum_{q \in \mcal{P}} \frac{\sigma_\beta \lp p, q \rp  \rho  \lp q \rp m_\beta \lp q \rp}{\sum_{r \in \mcal{P}} \sigma_\beta \lp r, q \rp \rho \lp r \rp} + m_{\beta} \lp p \rp, \label{simplify_beta} \\
    \Gamma \lp p \rp &= \sum_{q\in \mcal{P}} \frac{\sigma_\gamma \lp p, q \rp  \rho  \lp q \rp m_\gamma \lp q \rp}{\sum_{r \in \mcal{P}} \sigma_\gamma \lp r, q \rp \rho \lp r \rp} + m_{\gamma} \lp p \rp  . \label{simplify_gamma}
\end{align}

Note that the form of (\ref{beta_simple}) and (\ref{gamma_simple}) are possible due to the model formulation on the weighted graph. 

Plugging in all the components into (\ref{rate_eq}), with the procedure similar as before, we have  

\begin{align}
    &\quad - {\delta_{k, m_\alpha \lp p \rp} \rho \lp p \rp} \nonumber\\
    & = \lp A \lp p \rp + \frac{c_\beta}{c_\alpha} B \lp p \rp \rp n_{k-1} \lp p \rp +  \frac{c_\gamma}{c_\alpha} \Gamma \lp p \rp n_{k+1} \lp p \rp 
    \nonumber \\
     & \quad + \lp -  A \lp p \rp - \frac{c_\beta}{c_\alpha} B \lp p \rp -\frac{c_\gamma}{c_\alpha} \Gamma \lp p \rp -1  \rp n_{k} \lp p \rp
    ,
     \label{simple_solution}
\end{align}
which is a second-order linear recurrence relation. Since homogeneous linear recurrence relation is easily solvable given numerical value, we skip the closed form solution here.

\section{Case Studies: Extraversion and Agreeableness}
\label{modeling}

In this section, we consider how to practically model the effect of personality on friendship networks. We treat the effect of each traits separately in the following subsections. 

\subsection{Modeling the Effects of Extraversion}
\label{ext}
In this section, we discuss the modeling with regard to one of the most dominant personality trait, extraversion. 

We specify the components of the modeling as in Table \ref{tab:ext}, which are consistent with the results summarized by Harris and Vazire \cite{harris2016friendship}: extraverts are more attractive at zero acquaintance; extraverts are more likely to engage in social interactions which provide them the opportunity to meet new friends.
\begin{table}[!t]
    \centering
    \caption{Modeling Specification}
    \label{tab:ext}
    \begin{tabular}{ll}
        \hline 
        \\[-0.7em]
        \multicolumn{2}{l}{Space and Distribution for Personality}\\[0.3em]
        $\mcal{P} = \lb -1, 1 \rb$ & \\
        $\rho \lp p \rp = \frac{1}{2}, \forall p \in \mcal{P}$ & \\[0.3em]
        \hline 
        \\[-0.7em]
        \multicolumn{2}{l}{{Modeling the Effects of Extraversion}} \\[0.3em]
        $\pi_\alpha \lp \lp p, k \rp, \lp q, l \rp \rp = c_0 p + c_1$ 
        & $c_0>0 , c_1 \geq c_0$  \\
        $\pi_\beta \lp \lp p, k \rp, \lp q, l \rp \rp = c_2 $
        & $c_2 \geq 0$ \\
        $\pi_\gamma \lp \lp p, k \rp, \lp q, l \rp \rp = c_3$ 
        & $c_3 \geq 0$ \\
        $m_\alpha \lp p \rp = c_4$ 
        & $c_4 > 0$ \\
        $m_\beta \lp p \rp = c_5 p + c_6$ 
        & $c_5>0, c_6\geq c_5$ \\
        $m_\gamma \lp p \rp = c_7$ 
        & $c_7 \geq 0$ \\[0.3em]
        \hline 
        \\[-0.7em]
        \multicolumn{2}{l}{{Modeling the Effects of Agreeableness}} \\[0.3em]
        $\pi_\alpha \lp \lp p, k \rp, \lp q, l \rp \rp = c_0$
        & $c_0 \geq 0$  \\
        $\pi_\beta \lp \lp p, k \rp, \lp q, l \rp \rp = c_1 p + c_2$
        & $c_1 > 0, c_2 \geq c_1$ \\
        $\pi_\gamma \lp \lp p, k \rp, \lp q, l \rp \rp = c_3 p + c_4$
        & $c_3 < 0, c_4 \geq -c_3$ \\
        $m_\alpha \lp p \rp = c_5$
        & $c_5 > 0$\\
        $m_\beta \lp p \rp = c_6$
        & $c_6>0$ \\
        $m_\gamma \lp p \rp= c_7 p + c_8$
        & $c_7<0, c_8 \geq -c_7$ \\[0.3em]
        \hline
    \end{tabular}
\end{table}

Plugging into (\ref{simplify_alpha}), (\ref{simplify_beta}), and (\ref{simplify_gamma}), we can get the expression of $A \lp p \rp$, $B \lp p \rp$, and $\Gamma \lp p \rp$ to solve (\ref{simple_solution}). 
Since the closed form solution is formidable, we consider a special case for $c_\beta = c_\gamma = 0$. With (\ref{simplest-solution}), we have the degree-personality distribution as 

\begin{align}
    n_k \lp p \rp = \frac{c_1}{ c_1 + c_4 c_0 p + c_4 c_1} \lp \frac{ c_4 c_0 p + c_4 c_1}{ c_1 + c_4 c_0 p + c_4 c_1}  \rp^{k-c_4}.
    \label{dist_ext}
\end{align}

As shown in (\ref{conditional-exp_simplest}), the degree expectation conditioned on personality is 

\begin{align}
    \mb{E} \lb k \mv p \rb = \frac{c_4 \lp c_0 p + c_1 \rp}{ c_1} + c_4.
    \label{expected-degree_extra}
\end{align}

Computing the derivative of (\ref{expected-degree_extra}) , we have

\begin{align}
    \frac{\partial}{\partial p}  \mb{E} \lb k \mv p \rb = \frac{c_4 c_0}{ c_1} > 0. 
\end{align}

Hence, in the generated network, the expected degree is increased with the level of extraversion, which is consistent with many empirical results found \cite{pollet2011extraverts, feiler2015popularity}. 

\subsection{Modeling the Effects of Agreeableness}
\label{agr}

In this section, we consider another significant dimension in friendship, the agreeableness.  

Similar as in Section \ref{ext}, we specify the modeling components as in Table \ref{tab:ext}, which is consistent with the results summarized in \cite{harris2016friendship}: agreeable people are better liked after interacting with others; friendship with agreeable people participating is more satisfying; agreeable people conflict less and are better skilled at handling conflicts. 

Similarly, with (\ref{simplify_alpha}), (\ref{simplify_beta}), and (\ref{simplify_gamma}), we have expression of $A \lp p \rp$, $B \lp p \rp$, and $\Gamma \lp p \rp$, 
enabling us to solve (\ref{simple_solution}) numerically. 

To gain further insight, we consider a special case such that $c_\gamma = 0$. Under such condition, (\ref{simple_solution}) will be reduced to first-order form and easily solvable.
The solution is 

\begin{align}
    &n_k \lp p \rp \nonumber \\
    &= \frac{\rho \lp p \rp}{1 + A \lp p \rp + \frac{c_\beta}{c_\alpha} B \lp p \rp} \lp \frac{A \lp p \rp  + \frac{c_\beta}{c_\alpha} B \lp p \rp}{1 + A \lp p \rp  + \frac{c_\beta}{c_\alpha} B \lp p \rp} \rp^{k-m_\alpha \lp p \rp}.
    \label{dist_agr}
\end{align}

And the expected degree conditioned on personality is 

\begin{align}
    \mb{E} \lb k|p \rb &= A \lp p \rp + \frac{c_\beta}{c_\alpha} B \lp p \rp + m_\alpha \lp p \rp \nonumber \\
    &= \frac{c_\beta c_6 c_1}{ c_\alpha c_2}p + 2 \lp \frac{c_\beta}{c_\alpha} c_6 + c_5 \rp,
    \label{cond-exp_agre}
\end{align}
by substituting the specified components. 

Computing the derivative of (\ref{cond-exp_agre}), we have 

\begin{align}
    \frac{\partial}{\partial p}  \mb{E} \lb k \mv p \rb = \frac{c_\beta c_6 c_1}{ c_\alpha c_2} > 0.
\end{align}
That is, agreeable people have larger expected degree than the disagreeable, consistent with findings on the relationship between agreeableness and social networks \cite{zhu2013pathways,  wagner2014belongs}.
\section{Model Evaluation}
\label{validation}
Following our modeling and analysis in Section \ref{modeling}, we validate and discuss the implication of our case studies in this section. 

\subsection{Numerical Simulations}

To validate our analytically derived results in Section \ref{ext} and \ref{agr}, we simulate our models numerically. 

The simulation programs are implemented with Python programming language 
and network manipulating package \textit{graph-tool} \cite{peixoto_graph-tool_2014}, along with other packages for numerical computation and visualization 
.

To enabling the numerical simulation, all the abstract components described in Section \ref{ext} and \ref{agr} have to be substitute with numerical values. For all the simulation, models are initialized as a random graph with $N_0 = 15$ and $L_0 = 30$. For modeling-specific parameters and components, we summarize the specifications for simulation in Table \ref{tab:specification}. 


\begin{table}[!t]
    \centering
    \caption{specification for numerical simulations}
    \label{tab:specification}
    \begin{tabular}{c||c|c|c|c|c|c}
        Section & $\pi_\alpha $ & $\pi_\beta$ & $\pi_\gamma$ & $m_\alpha$ & $m_\beta$ & $ m_\gamma$  \\
         \hline
         \hline
        \ref{simul_ext} & $p+1$ & $2$ & $2$ & $10$ & $3p + 3$ & $3$ \\
        \hline
        \ref{simul_agr} & $ 1 $ & $ p+1 $ & $ -p+1 $ & $ 10 $ & $ 3 $ & $ -2p+2 $ 
    \end{tabular}
\end{table}

Each simulation takes $T=10000$ rounds; for each modeling, we run simulations for $10$ times in order to investigate the expected behavior of the models. We examine the estimated degree-personality joint distribution and degree expectation conditioned on personality. For estimating distribution, all the degree-personality pairs collected are used, with Gaussian kernel density estimation. For degree expectation estimation, we apply running average along the personality axis to compute the degree expectation conditioned on personality. 

\subsection{Simulating the Effect of Extraversion}
\label{simul_ext}
To validate our modeling for extraversion in Section \ref{ext}, we outline our simulation results below, with $\lp c_\alpha, c_\beta \rp = \lp 1, 0 \rp$. 

\begin{figure}[!t]
    \centering
    \subfloat[Estimated by simulation]{%
       \includegraphics[width=0.8\linewidth]{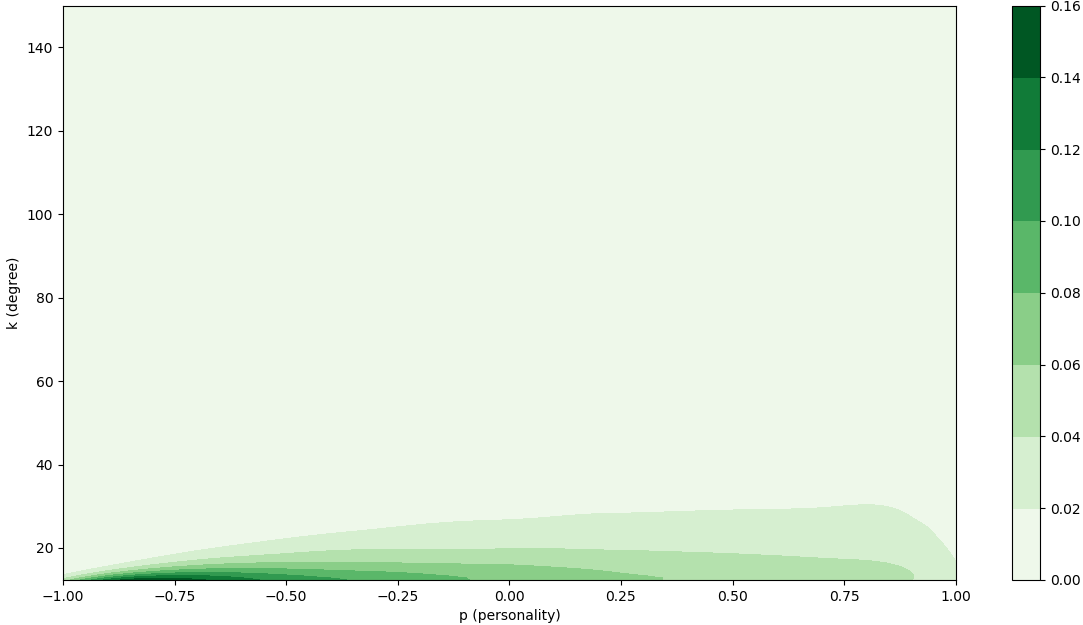}}
    \\
    \subfloat[Predicted by derivation]{%
        \includegraphics[width=0.8\linewidth]{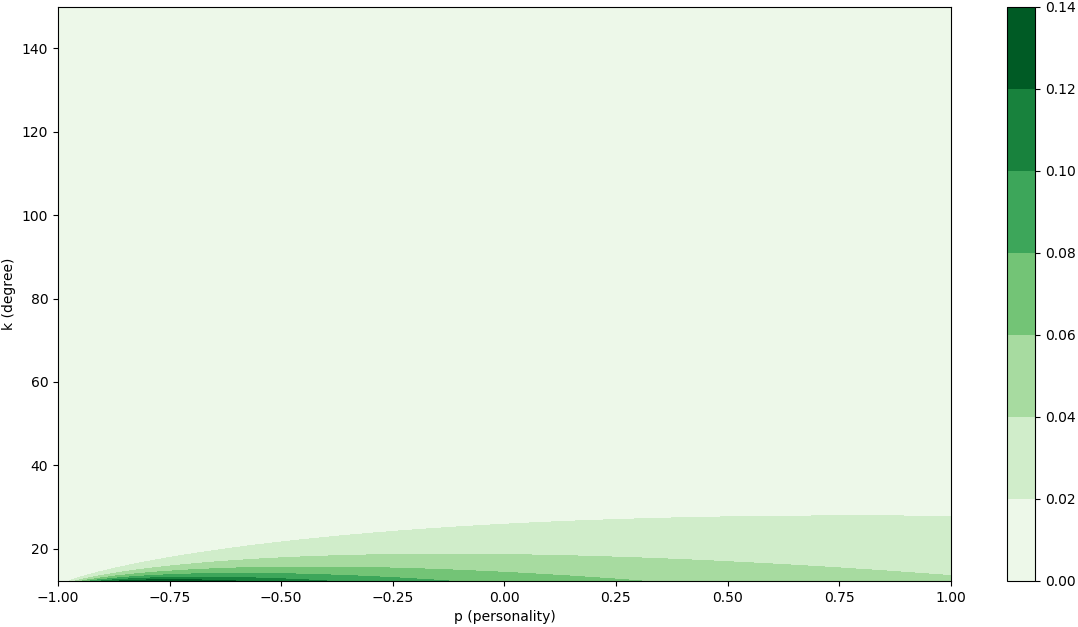}}
    \caption{Estimated and analytically derived degree-personality density.}
    \label{fig:ext_density}
\end{figure}

For degree-personality joint distribution, we compare the estimated density and the derived distribution in Fig. \ref{fig:ext_density}. The distribution is analytically derived in (\ref{dist_ext}), under $c_\beta = c_\gamma =0$. Neglecting the margin, it can be seen that the simulation result is approximately aligned with the derived analytical results.

\begin{figure}[!t]
    \centering
    \includegraphics[width = 0.8\linewidth]{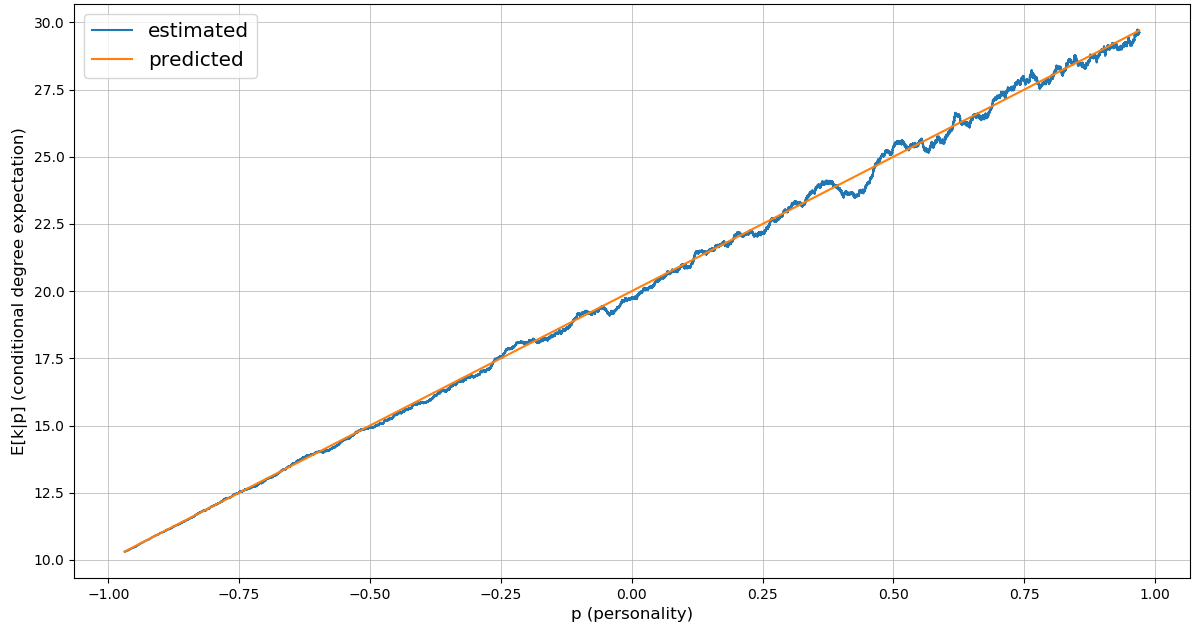}
    \caption{The estimated degree expectation conditioned on personality along with the derived analytical prediction. The curve is estimated by taking running average with window size $ W = 3000$.}
    \label{fig:ext_degree_expectation}
\end{figure}

On the other hand, the estimated degree expectation conditioned on personality is plotted in Fig. \ref{fig:ext_degree_expectation}, along with the analytical prediction derived in (\ref{expected-degree_extra}). 

As visually illustrated in Fig. \ref{fig:ext_degree_expectation} and claimed in Section \ref{ext}, in our model, agents' expected degrees are increasing with their levels of extraversion. This result is consistent with empirical evidences \cite{pollet2011extraverts, feiler2015popularity}. 

\subsection{Simulating the Effect of Agreeableness}
\label{simul_agr}
Similarly, we illustrate our simulation result to validate how we model the effect of agreeableness in Section \ref{agr}, with $\lp c_\alpha, c_\beta \rp = \lp 0.4, 0.6 \rp$.

\begin{figure}[!t]
    \centering
    \subfloat[Estimated by simulation]{%
       \includegraphics[width=0.8\linewidth]{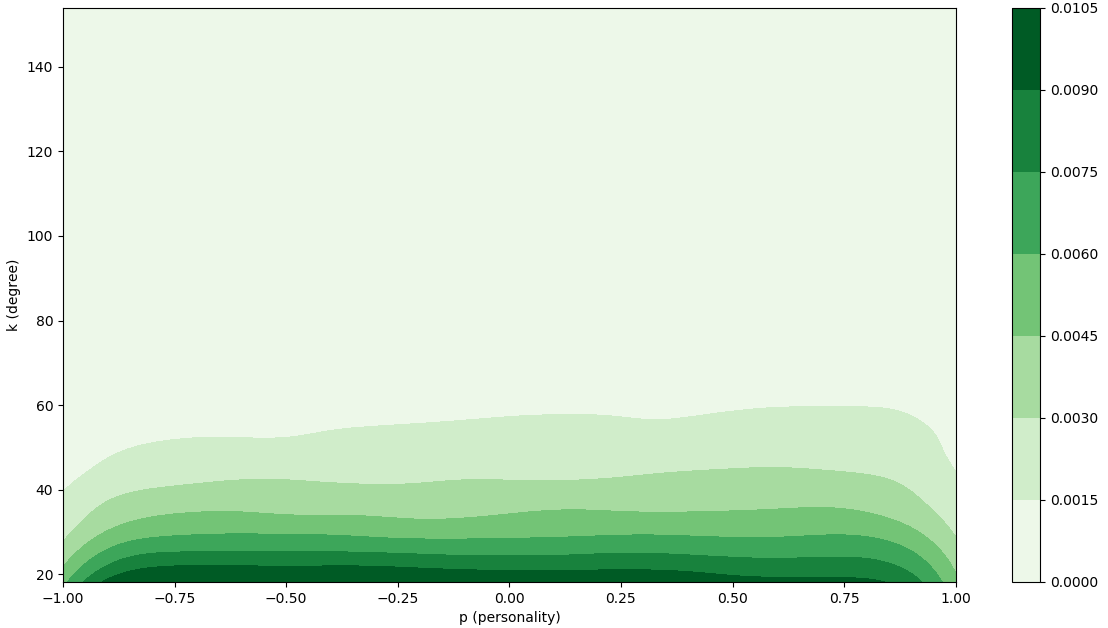}}
    \\
    \subfloat[Predicted by derivation]{%
        \includegraphics[width=0.8\linewidth]{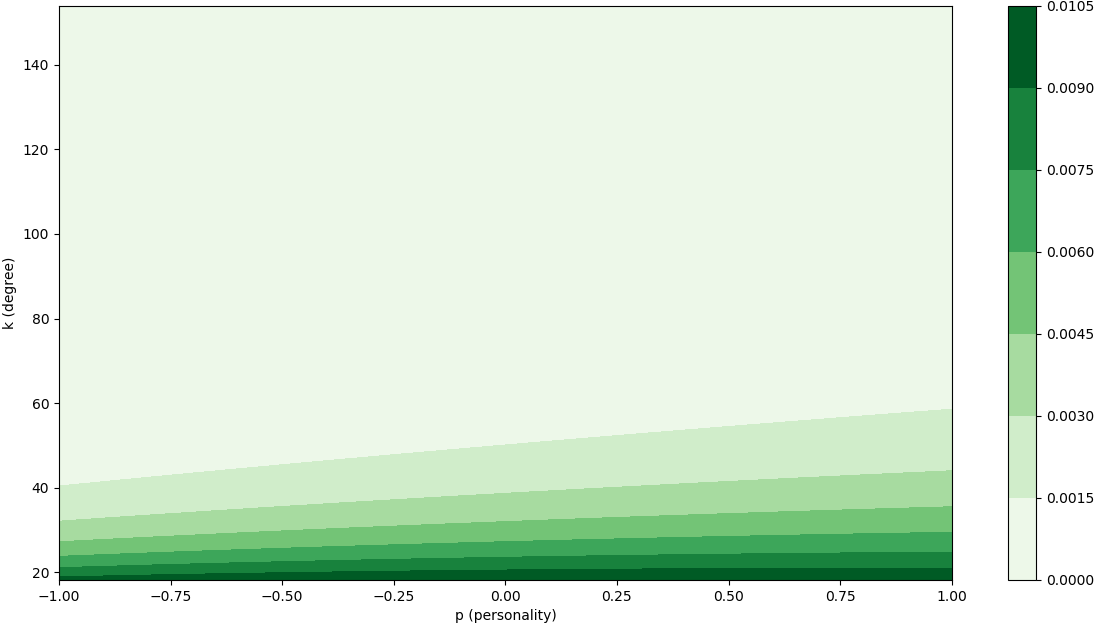}}
    \caption{Estimated and analytically derived degree-personality density.}
    \label{fig:agr_density}
\end{figure}

\begin{figure}[!t]
    \centering
    \includegraphics[width = 0.8\linewidth]{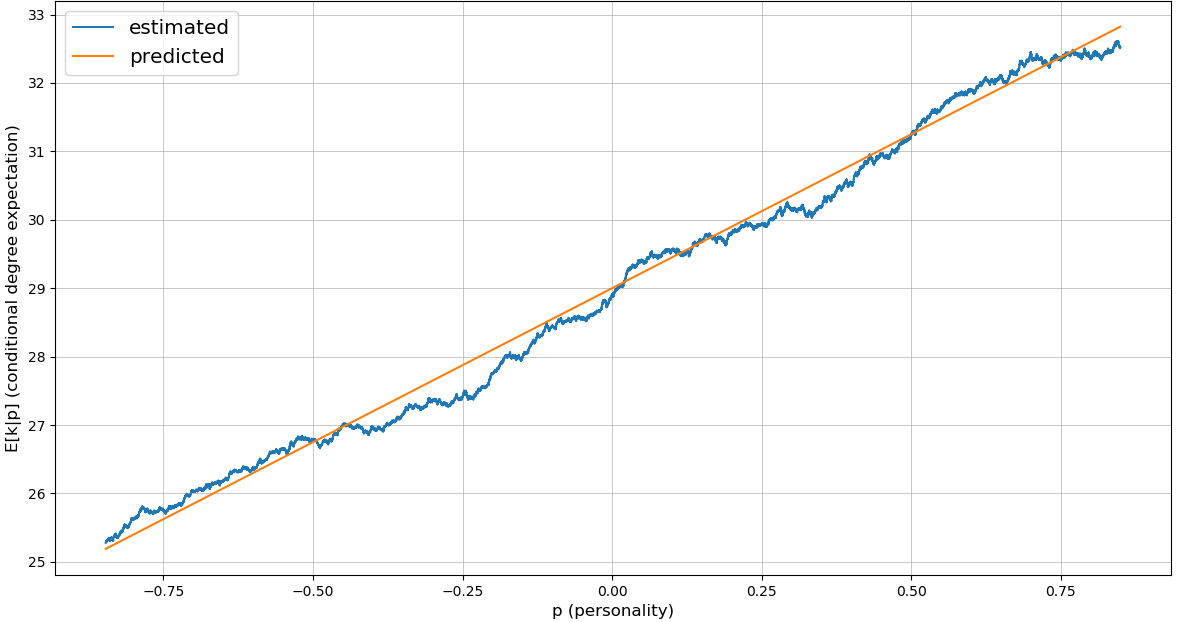}
    \caption{The estimated degree expectation conditioned on personality along with the derived analytical prediction. The curve is estimated by taking running average with window size $ W = 6000$.}
    \label{fig:agr_degree_expectation}
\end{figure}

The degree-personality joint distribution and expected degree conditioned on personality are shown in Fig. \ref{fig:agr_density} and Fig. \ref{fig:agr_degree_expectation}, respectively, both show their consistency with analytical prediction made in (\ref{dist_agr}) and (\ref{cond-exp_agre}).  

As visually illustrated in Fig. \ref{fig:agr_degree_expectation}, the degree expectation also tend to arise with agents' levels of agreeableness. Although the effect size might not be large compared to extraversion, this tendency was indeed found in some empirical researches \cite{zhu2013pathways, wagner2014belongs}.

\section{Discussions}
\label{discussion}
In this sections, we discuss how our results are consistent with existing literature, and also the limitation and potential further investigation of this study. 

\subsection{Consistency with Empirical Research}

We have provided short examples on using our proposed model to study how the effect of personality on the friendship developments lead to the effect of personality on positions in networks. We also predicted that both extraversion and agreeableness are positively correlated with degree in friendship networks in Section \ref{modeling} and \ref{validation}.

There have been lots of research discussing how the personality and network structure or position are correlated \cite{selden2018review}. For example, Table \ref{tab:emp} summarizes some prior works showing that extraversion and agreeableness are positively correlated with degree, which is captured by the prediction made by our models proposed.

\begin{table}[!t]
    \centering
    \caption{Summarized Empirical Studies  }
    \label{tab:emp}
    \begin{tabular}{c||c|c|c}
        Study & Results$^\mrm{a}$ & Relationship Type & Sample$^\mrm{b}$  \\
        \hline
        \hline 
        \cite{feiler2015popularity} & E+ & Friends & Graduate Students (US) \\
        \hline 
        \cite{selfhout2010emerging} & E+, A+ & Friends & College Students (NL) \\
        \hline
        \cite{pollet2011extraverts} & E+ & Relatives and friends & Adults (NL)  \\
        \hline
        \cite{zhu2013pathways} & E+, A+ & Social support & College Students (US) \\
        \hline 
        \cite{wagner2014belongs} & E+, A+ & Close others & High school (DE) \\
        \hline 
        \cite{kalish2006psychological} & E+ & Important contacts & College Students (IL) \\
        \hline 
        \cite{schulte2012coevolution} & E+, A+ & Friends, advice & Young adults (US) \\
        \hline 
        \multicolumn{4}{l}{$^\mrm{a}$`E+' and  `A+' indicates that degree is found to be positively correlated } \\
        \multicolumn{4}{l}{$\,\,$ with extraversion and agreeableness, respectively.} \\
        \multicolumn{4}{l}{$^\mrm{b}$countries represented by ISO 2 letter code. }
    \end{tabular}
\end{table}

While there are research endeavors showing that extraversion and agreeableness are not positively correlated with degree \cite{klein2004they, liu2010they, gloor2011towards}, most of them were not concerned with friendship networks, but dealing with other kinds of social networks instead.
Since our modeling is based on the interdependence between friendship and personality, it is possible that our modeling in Section \ref{modeling} cannot predict how personality and network structure are correlated outside the scope of friendship networks. 

It can be concluded that our modeling indeed captures how dependence of friendship development and personality leads to the meso-level patterns in social networks, which are supported by the empirical researches listed in Table \ref{tab:emp}. 

\subsection{Outlooks}

In Section \ref{model}, 
we analyze the model behavior when the model specification is satisfied with certain constraint. An analytical solution with regard to the most general model will be more desirable, which might be a promising future direction. 

In Section \ref{modeling}, we specified the model components using linear and constant functions consistent with existing results. It is desirable to further estimate such functions empirically by well-designed psychological studies. 

The results of our modeling were validated qualitatively by existing studies (summarized in Table \ref{tab:emp}); if more data about personality and friendship is available in the future, it is possible to further evaluate the prediction. On the other hand, we only considered the correlation between degree and personality when evaluating our results in Section \ref{modeling}. A potential next step is to take higher-order network measures into consideration
\section{Conclusions}
\label{conclusion}
In this study, we propose the first mathematical model of human friendship networks with personality. Preliminary analytical results were provided and some practical modeling scenarios were considered, with both analytical and simulation results. By using the proposed model, we show that the effect of personality on friendship development can aggregate to the effect of personality on friendship network structure. We hope this study will stimulate more research to investigate about the interplay of personality and friendship, with perspectives from network science specifically.

\section*{Acknowledgment}


H.C. Liou thanks the advice from Prof. Jen-Ho Chang during the early stages of this research. This work was supported by College Student Research grant number 108-2813-C-002 -036 -E from Ministry of Science and Technology of Taiwan.






{\small
\bibliographystyle{IEEEtran}
\bibliography{ref.bib}

\begin{thebibliography}{10}
\providecommand{\url}[1]{#1}
\csname url@samestyle\endcsname
\providecommand{\newblock}{\relax}
\providecommand{\bibinfo}[2]{#2}
\providecommand{\BIBentrySTDinterwordspacing}{\spaceskip=0pt\relax}
\providecommand{\BIBentryALTinterwordstretchfactor}{4}
\providecommand{\BIBentryALTinterwordspacing}{\spaceskip=\fontdimen2\font plus
\BIBentryALTinterwordstretchfactor\fontdimen3\font minus
  \fontdimen4\font\relax}
\providecommand{\BIBforeignlanguage}[2]{{%
\expandafter\ifx\csname l@#1\endcsname\relax
\typeout{** WARNING: IEEEtran.bst: No hyphenation pattern has been}%
\typeout{** loaded for the language `#1'. Using the pattern for}%
\typeout{** the default language instead.}%
\else
\language=\csname l@#1\endcsname
\fi
#2}}
\providecommand{\BIBdecl}{\relax}
\BIBdecl

\bibitem{burt2013social}
R.~S. Burt, M.~Kilduff, and S.~Tasselli, ``Social network analysis: Foundations
  and frontiers on advantage,'' \emph{Annual review of psychology}, vol.~64,
  pp. 527--547, 2013.

\bibitem{selden2018review}
M.~Selden and A.~S. Goodie, ``Review of the effects of five factor model
  personality traits on network structures and perceptions of structure,''
  \emph{Social Networks}, vol.~52, pp. 81--99, 2018.

\bibitem{fang2015integrating}
R.~Fang, B.~Landis, Z.~Zhang, M.~H. Anderson, J.~D. Shaw, and M.~Kilduff,
  ``Integrating personality and social networks: A meta-analysis of
  personality, network position, and work outcomes in organizations,''
  \emph{Organization Science}, vol.~26, no.~4, pp. 1243--1260, 2015.

\bibitem{feiler2015popularity}
D.~C. Feiler and A.~M. Kleinbaum, ``Popularity, similarity, and the network
  extraversion bias,'' \emph{Psychological science}, vol.~26, no.~5, pp.
  593--603, 2015.

\bibitem{selfhout2010emerging}
M.~Selfhout, W.~Burk, S.~Branje, J.~Denissen, M.~Van~Aken, and W.~Meeus,
  ``Emerging late adolescent friendship networks and big five personality
  traits: A social network approach,'' \emph{Journal of personality}, vol.~78,
  no.~2, pp. 509--538, 2010.

\bibitem{pollet2011extraverts}
T.~V. Pollet, S.~G. Roberts, and R.~I. Dunbar, ``Extraverts have larger social
  network layers,'' \emph{Journal of Individual Differences}, 2011.

\bibitem{zhu2013pathways}
X.~Zhu, S.~E. Woo, C.~Porter, and M.~Brzezinski, ``Pathways to happiness: From
  personality to social networks and perceived support,'' \emph{Social
  networks}, vol.~35, no.~3, pp. 382--393, 2013.

\bibitem{wagner2014belongs}
J.~Wagner, O.~L{\"u}dtke, B.~W. Roberts, and U.~Trautwein, ``Who belongs to me?
  social relationship and personality characteristics in the transition to
  young adulthood,'' \emph{European Journal of Personality}, vol.~28, no.~6,
  pp. 586--603, 2014.

\bibitem{kalish2006psychological}
Y.~Kalish and G.~Robins, ``Psychological predispositions and network structure:
  The relationship between individual predispositions, structural holes and
  network closure,'' \emph{Social networks}, vol.~28, no.~1, pp. 56--84, 2006.

\bibitem{schulte2012coevolution}
M.~Schulte, N.~A. Cohen, and K.~J. Klein, ``The coevolution of network ties and
  perceptions of team psychological safety,'' \emph{Organization Science},
  vol.~23, no.~2, pp. 564--581, 2012.

\bibitem{singer2009agent}
H.~M. Singer, I.~Singer, and H.~J. Herrmann, ``Agent-based model for friendship
  in social networks,'' \emph{Physical Review E}, vol.~80, no.~2, p. 026113,
  2009.

\bibitem{currarini2009economic}
S.~Currarini, M.~O. Jackson, and P.~Pin, ``An economic model of friendship:
  Homophily, minorities, and segregation,'' \emph{Econometrica}, vol.~77,
  no.~4, pp. 1003--1045, 2009.

\bibitem{ilany2016personality}
A.~Ilany and E.~Ak{\c{c}}ay, ``Personality and social networks: A generative
  model approach,'' \emph{Integrative and comparative biology}, vol.~56, no.~6,
  pp. 1197--1205, 2016.

\bibitem{lusseau2004identifying}
D.~Lusseau and M.~E. Newman, ``Identifying the role that animals play in their
  social networks,'' \emph{Proceedings of the Royal Society of London. Series
  B: Biological Sciences}, vol. 271, no. suppl\_6, pp. S477--S481, 2004.

\bibitem{mccrae1992introduction}
R.~R. McCrae and O.~P. John, ``An introduction to the five-factor model and its
  applications,'' \emph{Journal of personality}, vol.~60, no.~2, pp. 175--215,
  1992.

\bibitem{harris2016friendship}
K.~Harris and S.~Vazire, ``On friendship development and the big five
  personality traits,'' \emph{Social and Personality Psychology Compass},
  vol.~10, no.~11, pp. 647--667, 2016.

\bibitem{cervone2015personality}
D.~Cervone and L.~A. Pervin, \emph{Personality: Theory and research}.\hskip 1em
  plus 0.5em minus 0.4em\relax John Wiley \& Sons, 2015.

\bibitem{costa1995domains}
P.~T. Costa~Jr and R.~R. McCrae, ``Domains and facets: Hierarchical personality
  assessment using the revised neo personality inventory,'' \emph{Journal of
  personality assessment}, vol.~64, no.~1, pp. 21--50, 1995.

\bibitem{mcadams2015art}
D.~P. McAdams, \emph{The art and science of personality development}.\hskip 1em
  plus 0.5em minus 0.4em\relax Guilford Publications, 2015.

\bibitem{dunbar2018anatomy}
R.~I. Dunbar, ``The anatomy of friendship,'' \emph{Trends in cognitive
  sciences}, vol.~22, no.~1, pp. 32--51, 2018.

\bibitem{adams1994integrative}
R.~G. Adams and R.~Blieszner, ``An integrative conceptual framework for
  friendship research,'' \emph{Journal of Social and Personal Relationships},
  vol.~11, no.~2, pp. 163--184, 1994.

\bibitem{cuperman2009big}
R.~Cuperman and W.~Ickes, ``Big five predictors of behavior and perceptions in
  initial dyadic interactions: Personality similarity helps extraverts and
  introverts, but hurts “disagreeables”.'' \emph{Journal of personality and
  social psychology}, vol.~97, no.~4, p. 667, 2009.

\bibitem{joiner1999depression}
T.~E. Joiner, G.~I. Metalsky, J.~Katz, and S.~R. Beach, ``Depression and
  excessive reassurance-seeking,'' \emph{Psychological Inquiry}, vol.~10,
  no.~3, pp. 269--278, 1999.

\bibitem{ferretti2012features}
L.~Ferretti, M.~Cortelezzi, B.~Yang, G.~Marmorini, and G.~Bianconi, ``Features
  and heterogeneities in growing network models,'' \emph{Physical Review E},
  vol.~85, no.~6, p. 066110, 2012.

\bibitem{back2011closer}
M.~D. Back, S.~C. Schmukle, and B.~Egloff, ``A closer look at first sight:
  Social relations lens model analysis of personality and interpersonal
  attraction at zero acquaintance,'' \emph{European Journal of Personality},
  vol.~25, no.~3, pp. 225--238, 2011.

\bibitem{peixoto_graph-tool_2014}
\BIBentryALTinterwordspacing
T.~P. Peixoto, ``The graph-tool python library,'' \emph{figshare}, 2014.
  [Online]. Available: \url{http://figshare.com/articles/graph_tool/1164194}
\BIBentrySTDinterwordspacing

\bibitem{klein2004they}
K.~J. Klein, B.-C. Lim, J.~L. Saltz, and D.~M. Mayer, ``How do they get there?
  an examination of the antecedents of centrality in team networks,''
  \emph{Academy of Management Journal}, vol.~47, no.~6, pp. 952--963, 2004.

\bibitem{liu2010they}
Y.~Liu and M.~Ipe, ``How do they become nodes? revisiting team member network
  centrality,'' \emph{The Journal of Psychology}, vol. 144, no.~3, pp.
  243--258, 2010.

\bibitem{gloor2011towards}
P.~A. Gloor, K.~Fischbach, H.~Fuehres, C.~Lassenius, T.~Niinim{\"a}ki, D.~O.
  Olguin, S.~Pentland, A.~Piri, and J.~Putzke, ``Towards “honest signals”
  of creativity--identifying personality characteristics through microscopic
  social network analysis,'' \emph{Procedia-Social and Behavioral Sciences},
  vol.~26, pp. 166--179, 2011.

\end{thebibliography}
}


\end{document}